# Fundamental Properties of Lagrangian Density for A Gravitational System and Their Derivations

**Fang-Pei Chen**
Department of Physics, Dalian University of Technology, Dalian 116024, China.
E-mail: chenfap@dlut.edu.cn

**Abstract**   Through the discussion of the fundamental properties of Lagrangian density for a gravitational system, the theoretical foundations of the modified Einstein's field equations

$$R^{\mu\nu} - \frac{1}{2} g^{\mu\nu} R - \lambda g^{\mu\nu} - D^{\mu\nu} = -8\pi G T^{\mu\nu}_{(M)}$$

and the Lorentz and Levi-Civita's conservation laws $\frac{\partial}{\partial x^\mu}(\sqrt{-g}\, T^{\mu\nu}_{(M)} + \sqrt{-g}\, T^{\mu\nu}_{(G)}) = 0$ are systematically studied. Our study confirms the view that they could be used as the premises to establish a new cosmology.

## 1.  Introduction

In the reference [1] a new theory of cosmology has been established which leads to the following distinct properties of cosmos: the energy of matter field might originate from the gravitational field; the big bang might not have occurred; the fields of the dark energy and some parts of the dark matter would not be matter fields but might be gravitational fields. These distinct properties are deduced from the following premises: 1), the Lagrangian density of pure gravitational field is supposed to be

$$\sqrt{-g}\, L_G(x) = \frac{\sqrt{-g}}{16\pi G}\left[ R(x) + 2\lambda + 2D(t) \right],$$   from which the modified Einstein's field equations

$$R^{\mu\nu} - \frac{1}{2} g^{\mu\nu} R - \lambda g^{\mu\nu} - D^{\mu\nu} = -8\pi G T^{\mu\nu}_{(M)} \qquad (1)$$

Can be derived; 2), we follow Lorentz and Levi-Civita to use $\sqrt{-g}\, T^{\mu\nu}_{(G)} \stackrel{def}{=} 2 \frac{\delta(\sqrt{-g}\, L_G)}{\delta g_{\mu\nu}}$ as the energy-momentum tensor density for the gravitational field, therefore the conservation laws of energy-momentum tensor density for a gravitational system including matter fields and gravitational fields are the Lorentz and Levi-Civita's conservation laws:

$$\frac{\partial}{\partial x^\mu}(\sqrt{-g}\, T^{\mu\nu}_{(M)} + \sqrt{-g}\, T^{\mu\nu}_{(G)}) = 0 \qquad (2)$$

and
$$T^{\mu\nu}_{(M)} + T^{\mu\nu}_{(G)} = 0 \qquad (3)$$

but not the Einstein's conservation laws:

$$\frac{\partial}{\partial x^\mu}(\sqrt{-g}\, T^{\mu\nu}_{(M)} + \sqrt{-g}\, t^{\mu\nu}_{(G)}) = 0.$$

Some people doubt that Ref.[1] contains fundamental mistakes and do not believe the above conclusions of Ref.[1]. The aim of this paper is to clear up this doubt. For this purpose we shall study firstly the fundamental properties of Lagrangian density for a gravitational system in the more general case and then explain the rationality for the Lagrangian density of a pure gravitational field

$$\sqrt{-g}\, L_G(x) = \frac{\sqrt{-g}}{16\pi G}[R(x) + 2\lambda + 2D(t)] \quad \text{and the modified Einstein's field equations}$$

$$R^{\mu\nu} - \frac{1}{2}g^{\mu\nu}R - \lambda g^{\mu\nu} - D^{\mu\nu} = -8\pi G\, T^{\mu\nu}_{(M)}$$

At the same time we shall expound the correctness of the Lorentz and Levi-Civita's conservation laws and compare the virtues and the defects between the new theory of cosmology and the prevalent big bang cosmology. These studies might help clearing up the doubt about the premises of the new theory of cosmology and the conclusions of Ref.[1].

## 2. The fundamental properties of Lagrangian density for a gravitational system in the more general case

In gravitational theories the action integral $I = \int \sqrt{-g(x)}\, L(x)\, d^4x$ is always used to study the rules of gravitation [2, 3, 4]. Where g(x) is the determinant of $g_{\mu\nu}(x)$; $\sqrt{-g(x)}\, L(x)$ is the total Lagrangian density of a gravitational system, it can be split into two parts:

$\sqrt{-g(x)}\, L(x) = \sqrt{-g(x)}\, L_M(x) + \sqrt{-g(x)}\, L_G(x)$ ; $\sqrt{-g(x)}\, L_G(x)$ is called gravitational Lagrangian density which is composed of gravitational fields only, $\sqrt{-g(x)}\, L_M(x)$ is called matter Lagrangian density which is composed of both matter fields and gravitational fields. Therefore $\sqrt{-(g)}\, L_G(x)$ describe only pure gravitational fields; but besides describing matter fields, $\sqrt{-g(x)}\, L_M(x)$ describe also the interactions between gravitational field and matter field.

In the General Relativity, the gravitational Lagrangian is denoted by

$$L_G(x) = \frac{1}{16\pi G} R(x) \qquad (4)$$

or
$$L_G(x) = \frac{1}{16\pi G}[R(x) + 2\lambda] \qquad (5)$$

the matter Lagrangian $L_M(x)$ is denoted by

$$L_M(x) = L_M[\psi(x); \psi_{|\mu}(x); h^i_{.\mu}(x)] \tag{6}$$

where $\psi(x)$ is the matter field, $\psi_{|\mu}(x)$ is the covariant derivative of $\psi(x)$:

$$\psi_{|\mu}(x) = \psi_{,\mu}(x) + \frac{1}{2} h_i^\lambda(x) h_{j\lambda,\mu}(x) \sigma^{ij} \psi(x) \quad [5] \tag{7}$$

Since the great majority of the fundamental fields for matter field are spinors, it is necessary to use tetrad field $h^i_{.\mu}(x)$ [5]. The metric field $g_{\mu\nu}(x)$ is expressed by $g_{\mu\nu}(x) = h^i_{.\mu}(x) h^j_{.\nu}(x) \eta_{ij}$ and we have

$$h_i^{.\mu}(x) = \eta_{ij} g_{\mu\nu}(x) h^j_{.\nu}(x) \quad ; \qquad h_{i\nu,\lambda}(x) = \frac{\partial}{\partial x^\lambda} h_{i\nu}(x) \quad ; \qquad \text{etc.}$$

In the new theory of cosmology established at Ref.[1], the matter Lagrangian $L_M(x)$ is also denoted by Eq.(6):

$$L_M(x) = L_M[\psi(x); \psi_{|\mu}(x); h^i_{.\mu}(x)]$$

but the gravitational Lagrangian is denoted by

$$\sqrt{-g} L_G(x) = \frac{\sqrt{-g}}{16\pi G}[R(x) + 2\lambda + 2D(x)] \tag{8}$$

At here we suppose for the moment that $D(x)$ is a scalar function of {x}, from the cosmological principle it can be proved that $D(x) = D(t)$. Owing to Eq.(7) $\sqrt{-g(x)} L_M(x)$ can be denoted by the following functional form:

$$\sqrt{-g(x)} L_M(x) = \sqrt{-g(x)} L_M[\psi(x); \psi_{,\lambda}(x); h^i_{.\mu}(x); h^i_{.\mu,\lambda}(x)] \tag{9}$$

Since $$R = g^{\mu\nu} \left( \frac{\partial \Gamma^\lambda_{..\mu\nu}}{\partial x^\lambda} - \frac{\partial \Gamma^\lambda_{..\mu\lambda}}{\partial x^\nu} + \Gamma^\sigma_{..\mu\nu} \Gamma^\lambda_{..\lambda\sigma} - \Gamma^\sigma_{..\mu\lambda} \Gamma^\lambda_{..\nu\sigma} \right),$$

$$\Gamma^\lambda_{..\mu\nu} = \frac{1}{2} g^{\lambda\sigma} (g_{\sigma\mu,\nu} + g_{\sigma\nu,\mu} - g_{\mu\nu,\sigma})$$

and $h^i_{.\mu}(x)$ are the dynamical gravitational fields but $\lambda$ and $D(t)$ are only nondynamical constant or function, so $\sqrt{-g(x)} L_G(x)$ in Eqs.(4,5,8) all can be denoted by the functional form of the dynamical fields $h^i_{\cdot\mu}(x)$ and their derivatives:

$$\sqrt{-g(x)} L_G(x) = \sqrt{-g(x)} L_G [h^i_{\cdot\mu}(x); h^i_{\cdot\mu,\lambda}(x); h^i_{\cdot\mu,\lambda\sigma}(x)] \qquad (10)$$

Eq.(9) and Eq.(10) have summarized the general character of the General Relativity and the gravitational theory used at Ref.[1], consequently these two theories must all have the general properties deduced from Eq.(9) and Eq.(10) which we shall talk about in the following.

Symmetry exists universally in physical systems, one fundamental symmetry of a gravitational system is that the action integrals

$$I_M = \int \sqrt{-g(x)} L_M(x) d^4 x \qquad I_G = \int \sqrt{-g(x)} L_G(x) d^4 x \quad \text{and}$$

$$I = I_M + I_G = \int \sqrt{-g(x)} (L_M(x) + L_G(x)) d^4 x$$

satisfy $\delta I_M = 0$, $\delta I_G = 0$ and $\delta I = 0$ respectively under the following two simultaneous transformations [2,6]:

(1), the infinitesimal general coordinate transformation

$$x^\mu \to x'^\mu = x^\mu + \xi^\mu(x) \qquad (11)$$

(2), the local Lorentz transformation of tetrad frame

$$e_i(x) \to e'_i(x') = e_i(x) - \varepsilon^{mn}(x) \delta^j_m \eta_{ni} e_j(x) \qquad (12)$$

The symmetry (1) is precisely the symmetry of local space-time translations.

The sufficient condition of an action integral $I = \int \sqrt{-g(x)} L(x) d^4 x$ being $\delta I = 0$ under above transformations is [2,7]:

$$\delta_0(\sqrt{-g} L) + (\xi^\mu \sqrt{-g} L)_{,\mu} \equiv 0 \qquad (13)$$

where $\delta_0$ represent the variation at a fixed value of $x$. Evidently there are also the relations

$$\delta_0(\sqrt{-g} L_M) + (\xi^\mu \sqrt{-g} L_M)_{,\mu} \equiv 0 \text{ and } \delta_0(\sqrt{-g} L_G) + (\xi^\mu \sqrt{-g} L_G)_{,\mu} \equiv 0 \qquad (14)$$

If there exists only the symmetry (2), Eqs.(13,14) reduce to $\delta_0(\sqrt{-g} L) \equiv 0$, $\delta_0(\sqrt{-g} L_M) \equiv 0$ and $\delta_0(\sqrt{-g} L_G) \equiv 0$ respectively.

From Eq.(9) and Eq.(10) we have

$$\delta_0(\sqrt{-g}\, L_M) = \frac{\partial(\sqrt{-g}\, L_M)}{\partial \psi}\delta_0\psi + \frac{\partial(\sqrt{-g}\, L_M)}{\partial \psi_{,\lambda}}\delta_0\psi_{,\lambda} + \frac{\partial(\sqrt{-g}\, L_M)}{\partial h^i_{.\mu}}\delta_0 h^i_{.\mu}$$
$$+ \frac{\partial(\sqrt{-g}\, L_M)}{\partial h^i_{.\mu,\lambda}}\delta_0 h^i_{.\mu,\lambda} \tag{15}$$

$$\delta_0(\sqrt{-g}\, L_G) = \frac{\partial(\sqrt{-g}\, L_G)}{\partial h^i_{.\mu}}\delta_0 h^i_{.\mu} + \frac{\partial(\sqrt{-g}\, L_G)}{\partial h^i_{.\mu,\lambda}}\delta_0 h^i_{.\mu,\lambda} + \frac{\partial(\sqrt{-g}\, L_G)}{\partial h^i_{.\mu,\lambda\sigma}}\delta_0 h^i_{.\mu,\lambda\sigma} \tag{16}$$

As $\psi(x)$ is spinor and $h^i_{.\mu}(x)$ is both tetrad Lorentz vector and coordinate vector, under the infinitesimal general coordinate transformation and the local Lorentz transformation of tetrad frame, it is not difficult to derive the following induced variations [3]:

$$\delta_0\psi(x) = \frac{1}{2}\varepsilon^{mn}(x)\sigma_{mn}\psi(x) - \xi^\alpha(x)\psi_{,\alpha}(x) \tag{17}$$

$$\delta_0\psi_{,\lambda}(x) = \frac{1}{2}\varepsilon^{mn}(x)\sigma_{mn}\psi_{,\lambda}(x) - \frac{1}{2}\varepsilon^{mn}_{,\lambda}(x)\sigma_{mn}\psi(x) - \xi^\alpha(x)\psi_{,\alpha\lambda}(x)$$
$$- \xi^\alpha_{,\lambda}(x)\psi_{,\alpha}(x) \tag{18}$$

$$\delta_0 h^i_{.\mu}(x) = \varepsilon^{mn}(x)\delta^i_m \eta_{nj} h^j_{.\mu}(x) - \xi^\alpha_{,\mu}(x) h^i_{.\alpha}(x) - \xi^\alpha(x) h^i_{.\mu,\alpha}(x) \tag{19}$$

$$\delta_0 h^i_{.\mu,\lambda}(x) = \varepsilon^{mn}(x)\delta^i_m \eta_{nj} h^j_{.\mu,\lambda}(x) + \varepsilon^{mn}_{,\lambda}(x)\delta^i_m \eta_{nj} h^j_{.\mu}(x) - \xi^\alpha_{,\mu}(x) h^i_{.\alpha,\lambda}(x)$$
$$- \xi^\alpha_{,\mu\lambda}(x) h^i_{.\alpha}(x) - \xi^\alpha(x) h^i_{.\mu,\alpha\lambda}(x) - \xi^\alpha_{,\lambda}(x) h^i_{.\mu,\alpha} \tag{20}$$

$$\delta_0 h^i_{\cdot\mu,\lambda\sigma}(x) = \varepsilon^{mn}(x)\delta^i_m \eta_{nj} h^j_{\cdot\mu,\lambda\sigma}(x) + \varepsilon^{mn}_{,\sigma}\delta^i_m \eta_{nj} h^j_{\cdot\mu,\lambda}(x)$$
$$+ \varepsilon^{mn}_{,\lambda}\delta^i_m \eta_{nj} h^j_{\cdot\mu,\sigma}(x) + \varepsilon^{mn}_{,\lambda\sigma}\delta^i_m \eta_{nj} h^j_{\cdot\mu}(x) - \xi^\alpha_{,\mu}(x) h^i_{\cdot\alpha,\lambda\sigma}(x)$$
$$- \xi^\alpha_{,\mu\sigma}(x) h^i_{\cdot\alpha,\lambda}(x) - \xi^\alpha_{,\mu\lambda}(x) h^i_{\cdot\alpha,\sigma}(x) - \xi^\alpha_{,\mu\lambda\sigma}(x) h^i_{\cdot\alpha}(x) - \xi^\alpha(x) h^i_{\cdot\mu,\alpha\lambda\sigma}(x) \quad (21)$$
$$- \xi^\alpha_{,\sigma}(x) h^i_{\cdot\mu,\alpha\lambda}(x) - \xi^\alpha_{,\lambda}(x) h^i_{\cdot\mu,\alpha\sigma}(x) - \xi^\alpha_{,\lambda\sigma}(x) h^i_{\cdot\mu,\alpha}(x)$$

Putting Eqs.(17-21) into Eq.(15) and Eq.(16); using

$$\delta_0(\sqrt{-g}\,\Lambda) + (\xi^\mu \sqrt{-g}\,\Lambda)_{,\mu} \equiv 0, \text{ where } \Lambda = L_M \text{ or } \Lambda = L_G \text{ or } \Lambda = L_M + L_G;$$

owing to the independent arbitrariness of $\varepsilon^{mn}(x)$, $\varepsilon^{mn}_{,\lambda}(x)$, $\varepsilon^{mn}_{,\lambda\sigma}(x)$, $\xi^\alpha(x)$, $\xi^\alpha_{,\mu}(x)$,

$\xi^\alpha_{,\mu\lambda}(x)$ and $\xi^\alpha_{,\mu\lambda\sigma}(x)$, we obtain the following identities:

$$\frac{1}{2}\frac{\partial(\sqrt{-g}\,\Lambda)}{\partial \psi}\sigma_{mn}\psi + \frac{1}{2}\frac{\partial(\sqrt{-g}\,\Lambda)}{\partial \psi_{,\lambda}}\sigma_{mn}\psi_{,\lambda} + \frac{\partial(\sqrt{-g}\,\Lambda)}{\partial h^m_{\cdot\mu}}h_{n\mu}$$
$$+ \frac{\partial(\sqrt{-g}\,\Lambda)}{\partial h^m_{\cdot\mu,\lambda}}h_{n\mu,\lambda} = 0 \quad (22)$$

$$\frac{1}{2}\frac{\partial(\sqrt{-g}\,\Lambda)}{\partial \psi_{,\lambda}}\sigma_{mn}\psi + \frac{\partial(\sqrt{-g}\,\Lambda)}{\partial h^m_{\cdot\mu,\lambda}}h_{n\mu} + 2\frac{\partial(\sqrt{-g}\,\Lambda)}{\partial h^m_{\cdot\mu,\lambda\sigma}}\Gamma_{\cdot n\mu,\sigma} = 0 \quad (23)$$

$$\frac{\partial(\sqrt{-g}\,\Lambda)}{\partial h^m_{\cdot\mu,\lambda\sigma}}h_{n\mu} = \frac{\partial(\sqrt{-g}\,\Lambda)}{\partial h^n_{\cdot\mu,\lambda\sigma}}h_{m\mu} \quad (24)$$

$$\frac{\partial(\sqrt{-g}\,\Lambda)}{\partial\psi}\psi_{,\alpha}+\frac{\partial(\sqrt{-g}\,\Lambda)}{\partial\psi_{,\lambda}}\psi_{,\lambda\alpha}+\frac{\partial(\sqrt{-g}\,\Lambda)}{\partial h^{i}_{.\mu}}h^{i}_{.\mu,\alpha}$$
$$+\frac{\partial(\sqrt{-g}\,\Lambda)}{\partial h^{i}_{.\mu,\lambda}}h^{i}_{.\mu,\lambda\alpha}+\frac{\partial(\sqrt{-g}\,\Lambda)}{\partial h^{i}_{.\mu,\lambda\sigma}}h^{i}_{.\mu,\lambda\sigma\alpha}-(\sqrt{-g}\,\Lambda)_{,\alpha}=0 \tag{25}$$

$$\frac{\partial(\sqrt{-g}\,\Lambda)}{\partial\psi_{,\lambda}}\psi_{,\alpha}+\frac{\partial(\sqrt{-g}\,\Lambda)}{\partial h^{i}_{.\lambda}}h^{i}_{.\alpha}+\frac{\partial(\sqrt{-g}\,\Lambda)}{\partial h^{i}_{.\mu,\lambda}}h^{i}_{.\mu,\alpha}+\frac{\partial(\sqrt{-g}\,\Lambda)}{\partial h^{i}_{.\lambda,\mu}}h^{i}_{.\alpha,\mu}$$
$$+\frac{\partial(\sqrt{-g}\,\Lambda)}{\partial h^{i}_{.\lambda,\mu\sigma}}h^{i}_{.\alpha,\mu\sigma}+2\frac{\partial(\sqrt{-g}\,\Lambda)}{\partial h^{i}_{.\mu,\lambda\sigma}}h^{i}_{.\mu,\sigma\alpha}-\sqrt{-g}\,\Lambda\,\delta^{\lambda}_{\alpha}=0 \tag{26}$$

$$\frac{\partial(\sqrt{-g}\,\Lambda)}{\partial h^{i}_{.\mu,\lambda}}h^{i}_{.\alpha}+\frac{\partial(\sqrt{-g}\,\Lambda)}{\partial h^{i}_{.\mu,\lambda\sigma}}h^{i}_{.\alpha,\sigma}+\frac{\partial(\sqrt{-g}\,\Lambda)}{\partial h^{i}_{.\sigma,\lambda\mu}}h^{i}_{.\sigma,\alpha}-\frac{\partial}{\partial x^{\sigma}}(\frac{\partial(\sqrt{-g}\,\Lambda)}{\partial h^{i}_{.\mu,\lambda\sigma}})h^{i}_{.\alpha}$$
$$=-\frac{\partial}{\partial x^{\sigma}}(\frac{\partial(\sqrt{-g}\,\Lambda)}{\partial h^{i}_{.\mu,\lambda\sigma}}h^{i}_{.\alpha}) \tag{27}$$

$$\frac{\partial(\sqrt{-g}\,\Lambda)}{\partial h^{i}_{.\mu,\lambda\sigma}}h^{i}_{.\alpha}+\frac{\partial(\sqrt{-g}\,\Lambda)}{\partial h^{i}_{.\lambda\sigma,\mu}}h^{i}_{.\alpha}+\frac{\partial(\sqrt{-g}\,\Lambda)}{\partial h^{i}_{.\sigma,\mu\lambda}}h^{i}_{.\alpha}=0 \tag{28}$$

From Eq.(28) the another identity:
$$\frac{\partial^{3}}{\partial x^{\mu}\partial x^{\lambda}\partial x^{\sigma}}(\frac{\partial(\sqrt{-g}\,\Lambda)}{\partial h^{i}_{.\mu,\lambda\sigma}}h^{i}_{.\alpha})=0 \tag{29}$$

can be deduced.

Eqs.(25-28) stem from the symmetry of transformation Eq.(11), the conservation laws of energy--momentum tensor density for a gravitational system can be derived from these identities; we shall discuss the derivation in the following section. Eqs.(22-24) stem from the symmetry of transformation Eq.(12), the conservation laws of spin density for a gravitational system [2] can be derived from these identities; since the conservation laws of spin density for a gravitational system require to study specially, we shall not discuss the problem in this paper.

## 3. Equations of fields and Conservation laws of energy-momentum tensor density for a gravitational system

The equations of fields for a gravitational system can be derived from

$$\delta_0 I = \int \delta_0 ( \sqrt{-g(x)} L(x)) d^4 x = 0 \tag{30}$$

where $\sqrt{-g(x)} L(x) = \sqrt{-g(x)} L_M(x) + \sqrt{-g(x)} L_G(x)$, $\delta_0 I$ is the variation of $I$ correspponding to the variations of the dynamical field variable for the gravitational system at a fixed value of $x$. If the dynamical field variables of the gravitational system are $\psi(x), h^i_{.\mu}(x)$, then from Eq.(9) and Eq.(10)

$$\delta_0(\sqrt{-g} L) = \frac{\partial(\sqrt{-g} L)}{\partial \psi} \delta_0 \psi + \frac{\partial(\sqrt{-g} L)}{\partial \psi_{,\lambda}} \delta_0 \psi_{,\lambda} + \frac{\partial(\sqrt{-g} L)}{\partial h^i_{.\mu}} \delta_0 h^i_{.\mu}$$
$$+ \frac{\partial(\sqrt{-g} L)}{\partial h^i_{.\mu,\lambda}} \delta_0 h^i_{.\mu,\lambda} + \frac{\partial(\sqrt{-g} L)}{\partial h^i_{.\mu,\lambda\sigma}} \delta_0 h^i_{.\mu,\lambda\sigma} \tag{31}$$

where $\delta_0 \psi(x), \delta_0 h^i_{.\mu}(x)$ are arbitrary and independent variations, they may be or may not be symmetrical variations.

Substituing Eq.(31) into Eq.(30), using Gauss' theorem, and setting $\delta_0 \psi(x); \delta_0 h^i_{.\mu}(x)$ and their derivatives all equal to zero at the integration limits, we find

$$\left(\frac{\partial(\sqrt{-g} L)}{\partial \psi} - \frac{\partial}{\partial x^\lambda} \frac{\partial(\sqrt{-g} L)}{\partial \psi_{,\lambda}}\right) \delta_0 \psi$$
$$+ \left(\frac{\partial(\sqrt{-g} L)}{\partial h^i_{.\mu}} - \frac{\partial}{\partial x^\lambda} \frac{\partial(\sqrt{-g} L)}{\partial h^i_{.\mu,\lambda}} + \frac{\partial^2}{\partial x^\lambda \partial x^\sigma} \frac{\partial(\sqrt{-g} L)}{\partial h^i_{.\mu,\lambda\sigma}}\right) \delta_0 h^i_{.\mu} = 0 \tag{32}$$

Since $\psi(x), h^i_{.\mu}(x)$ are independent dynamical field variables, Eq. (32) is equivalent to the following two equations:

$$\frac{\delta(\sqrt{-g} L_M)}{\delta \psi} = \frac{\partial(\sqrt{-g} L_M)}{\partial \psi} - \frac{\partial}{\partial x^\mu} \frac{\partial(\sqrt{-g} L_M)}{\partial \psi_{,\mu}} = 0 \tag{33}$$

$$\frac{\delta(\sqrt{-g}\,L)}{\delta h^i_{.\mu}} = \frac{\partial(\sqrt{-g}\,L)}{\partial h^i_{.\mu}} - \frac{\partial}{\partial x^\lambda}\frac{\partial(\sqrt{-g}\,L)}{\partial h^i_{.\mu,\lambda}} + \frac{\partial^2}{\partial x^\lambda \partial x^\sigma}\frac{\partial(\sqrt{-g}\,L_G)}{\partial h^i_{.\mu,\lambda\sigma}} = 0 \tag{34}$$

Eq.(34) is always rewritten in the following form:

$$\begin{aligned}\frac{\delta(\sqrt{-g}\,L_G)}{\delta h^i_{.\mu}} &= \frac{\partial(\sqrt{-g}\,L_G)}{\partial h^i_{.\mu}} - \frac{\partial}{\partial x^\lambda}\frac{\partial(\sqrt{-g}\,L_G)}{\partial h^i_{.\mu,\lambda}} + \frac{\partial^2}{\partial x^\lambda \partial x^\sigma}\frac{\partial(\sqrt{-g}\,L_G)}{\partial h^i_{.\mu,\lambda\sigma}} \\ &= -\frac{\partial(\sqrt{-g}\,L_M)}{\partial h^i_{.\mu}} + \frac{\partial}{\partial x^\lambda}\frac{\partial(\sqrt{-g}\,L_M)}{\partial h^i_{.\mu,\lambda}} = -\frac{\delta(\sqrt{-g}\,L_M)}{\delta h^i_{.\mu}}\end{aligned} \tag{34'}$$

Eq.(33) is the equation of matter field; Eq.(34) or Eq.(34') are the equations of vierbein field $h^i_{.\mu}(x)$ which are gravitational fields.

It is well known that, in the special relativity, the conservation laws of energy-momentum tensor density for a physical system is originated from the action integral $I = \int \sqrt{-g(x)}\,L(x)d^4x$ of this physical system being invariant under space-time finite translations [8]. In relativistic theories of gravitation, there are no symmetry of space-time finite translations but have only the symmetry of local space-time translations $x^\mu \to x'^\mu = x^\mu + \xi^\mu(x)$, which is equivalent to the infinitesimal general coordinate transformation Eq.(11). In the following we shall use this local symmetry to deduce some identities which might be regarded as the conservation laws of energy-momentum for a gravitational system.

Eq.(25) can be transformed into

$$\begin{aligned}&\frac{\partial}{\partial x^\lambda}\Big(\sqrt{-g}\,\Lambda\,\delta^\lambda_\alpha - \frac{\partial(\sqrt{-g}\,\Lambda)}{\partial \psi_{,\lambda}}\psi_{,\alpha} - \frac{\partial(\sqrt{-g}\,\Lambda)}{\partial h^i_{.\mu,\lambda}}h^i_{.\mu,\alpha} - \frac{\partial(\sqrt{-g}\,\Lambda)}{\partial h^i_{.\mu,\lambda\sigma}}h^i_{.\mu,\sigma\alpha} \\ &+ \frac{\partial}{\partial x^\sigma}\Big(\frac{\partial(\sqrt{-g}\,\Lambda)}{\partial h^i_{.\mu,\lambda\sigma}}\Big)h^i_{.\mu,\alpha}\Big) = \Big(\frac{\partial(\sqrt{-g}\,\Lambda)}{\partial \psi} - \frac{\partial}{\partial x^\lambda}\Big(\frac{\partial(\sqrt{-g}\,\Lambda)}{\partial \psi_{,\lambda}}\Big)\Big)\psi_{,\alpha} \\ &+ \Big(\frac{\partial(\sqrt{-g}\,\Lambda)}{\partial h^i_{.\mu}} - \frac{\partial}{\partial x^\lambda}\Big(\frac{\partial(\sqrt{-g}\,\Lambda)}{\partial h^i_{.\mu,\lambda}}\Big) + \frac{\partial^2}{\partial x^\lambda \partial x^\sigma}\Big(\frac{\partial(\sqrt{-g}\,\Lambda)}{\partial h^i_{.\mu,\lambda\sigma}}\Big)\Big)h^i_{.\mu,\alpha}\end{aligned} \tag{35}$$

Utilizing Eq.(27), Eq.(26) can be transformed into

$$\sqrt{-g}\,\Lambda\,\delta^\lambda_\alpha - \frac{\partial(\sqrt{-g}\,\Lambda)}{\partial \psi_{,\lambda}}\psi_{,\alpha} - \frac{\partial(\sqrt{-g}\,\Lambda)}{\partial h^i_{.\mu,\lambda}}h^i_{.\mu,\alpha} - \frac{\partial(\sqrt{-g}\,\Lambda)}{\partial h^i_{.\mu,\lambda\sigma}}h^i_{.\mu,\sigma\alpha}$$

$$+ \frac{\partial}{\partial x^\sigma}\left(\frac{\partial(\sqrt{-g}\,\Lambda)}{\partial h^i_{.\mu,\lambda\sigma}}\right)h^i_{.\mu,\alpha} + \frac{\partial^2}{\partial x^\mu \partial x^\sigma}\left(\frac{\partial(\sqrt{-g}\,\Lambda)}{\partial h^i_{.\lambda,\mu\sigma}}h^i_{.\alpha}\right) \quad (36)$$

$$= \left(\frac{\partial(\sqrt{-g}\,\Lambda)}{\partial h^i_{.\lambda}} - \frac{\partial}{\partial x^\mu}\left(\frac{\partial(\sqrt{-g}\,\Lambda)}{\partial h^i_{.\lambda,\mu}}\right) + \frac{\partial^2}{\partial x^\mu \partial x^\sigma}\left(\frac{\partial(\sqrt{-g}\,\Lambda)}{\partial h^i_{.\lambda,\mu\sigma}}\right)\right)h^i_{.\alpha}$$

Let $\Lambda = L_M + L_G$ and use the equations of fields Eqs.(33,34), from Eq.(35) and Eq.(36) we get respectively:

$$\frac{\partial}{\partial x^\lambda}\left(\sqrt{-g}(L_M + L_G)\delta^\lambda_\alpha - \frac{\partial(\sqrt{-g}\,L_M)}{\partial \psi_{,\lambda}}\psi_{,\alpha} - \frac{\partial(\sqrt{-g}(L_M + L_G))}{\partial h^i_{.\mu,\lambda}}h^i_{.\mu,\alpha}\right.$$

$$\left. - \frac{\partial(\sqrt{-g}\,L_G)}{\partial h^i_{.\mu,\lambda\sigma}}h^i_{.\mu,\sigma\alpha} + \frac{\partial}{\partial x^\sigma}\left(\frac{\partial(\sqrt{-g}\,L_G)}{\partial h^i_{.\mu,\lambda\sigma}}\right)h^i_{.\mu,\alpha}\right) = 0 \quad (37)$$

and

$$\sqrt{-g}(L_M + L_G)\delta^\lambda_\alpha - \frac{\partial(\sqrt{-g}\,L_M)}{\partial \psi_{,\lambda}}\psi_{,\alpha} - \frac{\partial(\sqrt{-g}(L_M + L_G))}{\partial h^i_{.\mu,\lambda}}h^i_{.\mu,\alpha}$$

$$- \frac{\partial(\sqrt{-g}\,L_G)}{\partial h^i_{.\mu,\lambda\sigma}}h^i_{.\mu,\sigma\alpha} + \frac{\partial}{\partial x^\sigma}\left(\frac{\partial(\sqrt{-g}\,L_G)}{\partial h^i_{.\mu,\lambda\sigma}}\right)h^i_{.\mu,\alpha} + \frac{\partial^2}{\partial x^\mu \partial x^\sigma}\left(\frac{\partial(\sqrt{-g}\,L_G)}{\partial h^i_{.\mu,\lambda\sigma}}h^i_{.\alpha}\right) = 0 \quad (38)$$

Let $\Lambda = L_M$ and $\Lambda = L_G$, we get the further relations:

$$\frac{\partial}{\partial x^\lambda}\left(\sqrt{-g}\,L_M\,\delta^\lambda_\alpha - \frac{\partial(\sqrt{-g}\,L_M)}{\partial \psi_{,\lambda}}\psi_{,\alpha} - \frac{\partial(\sqrt{-g}\,L_M)}{\partial h^i_{.\mu,\lambda}}h^i_{.\mu,\alpha}\right) \quad (39)$$

$$= \left(\frac{\partial(\sqrt{-g}\,L_M)}{\partial h^i_{.\mu}} - \frac{\partial}{\partial x^\lambda}\left(\frac{\partial(\sqrt{-g}\,L_M)}{\partial h^i_{.\mu,\lambda}}\right)\right)h^i_{.\mu,\alpha}$$

$$\frac{\partial}{\partial x^\lambda}(\sqrt{-g}\,L_G\,\delta_\alpha^\lambda - \frac{\partial(\sqrt{-g}\,L_G)}{\partial h^i_{.\mu,\lambda}}h^i_{.\mu,\alpha} - \frac{\partial(\sqrt{-g}\,L_G)}{\partial h^i_{.\mu,\lambda\sigma}}h^i_{.\mu,\sigma\alpha}$$

$$+\frac{\partial}{\partial x^\sigma}(\frac{\partial(\sqrt{-g}\,L_G)}{\partial h^i_{.\mu,\lambda\sigma}})h^i_{.\mu,\alpha}) \qquad (40)$$

$$=(\frac{\partial(\sqrt{-g}\,L_G)}{\partial h^i_{.\mu}} - \frac{\partial}{\partial x^\lambda}(\frac{\partial(\sqrt{-g}\,L_G)}{\partial h^i_{.\mu,\lambda}}) + \frac{\partial^2}{\partial x^\lambda \partial x^\sigma}(\frac{\partial(\sqrt{-g}\,L_G)}{\partial h^i_{.\mu,\lambda\sigma}}))h^i_{.\mu,\alpha}$$

$$\sqrt{-g}\,L_M\,\delta_\alpha^\lambda - \frac{\partial(\sqrt{-g}\,L_M)}{\partial \psi_{,\lambda}}\psi_{,\alpha} - \frac{\partial(\sqrt{-g}\,L_M)}{\partial h^i_{.\mu,\lambda}}h^i_{.\mu,\alpha} \qquad (41)$$

$$=(\frac{\partial(\sqrt{-g}\,L_M)}{\partial h^i_{.\lambda}} - \frac{\partial}{\partial x^\mu}(\frac{\partial(\sqrt{-g}\,L_M)}{\partial h^i_{.\lambda,\mu}}))h^i_{.\alpha}$$

$$\sqrt{-g}\,L_G\,\delta_\alpha^\lambda - \frac{\partial(\sqrt{-g}\,L_G)}{\partial h^i_{.\mu,\lambda}}h^i_{.\mu,\alpha} - \frac{\partial(\sqrt{-g}\,L_G)}{\partial h^i_{.\mu,\lambda\sigma}}h^i_{.\mu,\sigma\alpha}$$

$$+\frac{\partial}{\partial x^\sigma}(\frac{\partial(\sqrt{-g}\,L_G)}{\partial h^i_{.\mu,\lambda\sigma}})h^i_{.\mu,\alpha} + \frac{\partial^2}{\partial x^\mu \partial x^\sigma}(\frac{\partial(\sqrt{-g}\,L_G)}{\partial h^i_{.\lambda,\mu\sigma}}h^i_{.\alpha}) \qquad (42)$$

$$=(\frac{\partial(\sqrt{-g}\,L_G)}{\partial h^i_{.\lambda}} - \frac{\partial}{\partial x^\mu}(\frac{\partial(\sqrt{-g}\,L_G)}{\partial h^i_{.\lambda,\mu}}) + \frac{\partial^2}{\partial x^\mu \partial x^\sigma}(\frac{\partial(\sqrt{-g}\,L_G)}{\partial h^i_{.\lambda,\mu\sigma}}))h^i_{.\alpha}$$

Rewrite Eq.(37) as

$$\frac{\partial}{\partial x^{\lambda}}(\sqrt{-g}\,L_M\,\delta^{\lambda}_{\alpha} - \frac{\partial(\sqrt{-g}\,L_M)}{\partial \psi_{,\lambda}}\psi_{,\alpha} - \frac{\partial(\sqrt{-g}\,L_M)}{\partial h^i_{.\mu,\lambda}}h^i_{.\mu,\alpha}$$

$$+\sqrt{-g}\,L_G\,\delta^{\lambda}_{\alpha} - \frac{\partial(\sqrt{-g}\,L_G)}{\partial h^i_{.\mu,\lambda}}h^i_{.\mu,\alpha} - \frac{\partial(\sqrt{-g}\,L_G)}{\partial h^i_{.\mu,\lambda\sigma}}h^i_{.\mu,\sigma\alpha} \qquad (37')$$

$$+\frac{\partial}{\partial x^{\sigma}}(\frac{\partial(\sqrt{-g}\,L_G)}{\partial h^i_{.\mu,\lambda\sigma}})h^i_{.\mu,\alpha}) = 0$$

This relation might be looked upon as a conservation laws of energy-momentum tensor density for gravitational system; owing to that in the special relativity $L_M\,\delta^{\lambda}_{\alpha} - \frac{\partial L_M}{\partial \psi_{,\lambda}}\psi_{,\alpha}$ is the energy-momentum tensor of matter field, $\sqrt{-g}\,L_M\,\delta^{\lambda}_{\alpha} - \frac{\partial(\sqrt{-g}\,L_M)}{\partial \psi_{,\lambda}}\psi_{,\alpha} - \frac{\partial(\sqrt{-g}\,L_M)}{\partial h^i_{.\mu,\lambda}}h^i_{.\mu,\alpha}$ should be interpreted as the energy-momentum tensor density of matter field, $\frac{\partial(\sqrt{-g}\,L_M)}{\partial h^i_{.\mu,\lambda}}h^i_{.\mu,\alpha}$ is the influence of gravitational field.

$$\sqrt{-g}\,L_G\,\delta^{\lambda}_{\alpha} - \frac{\partial(\sqrt{-g}\,L_G)}{\partial h^i_{.\mu,\lambda}}h^i_{.\mu,\alpha} - \frac{\partial(\sqrt{-g}\,L_G)}{\partial h^i_{.\mu,\lambda\sigma}}h^i_{.\mu,\sigma\alpha} + \frac{\partial}{\partial x^{\sigma}}(\frac{\partial(\sqrt{-g}\,L_G)}{\partial h^i_{.\mu,\lambda\sigma}})h^i_{.\mu,\alpha}$$ might be interpreted as the energy-momentum tensor density of pure gravitational field. Consequently the energy-momentum tensor density of matter field is always definded as

$$\sqrt{-g}\,T^{.\lambda}_{(M)i} \stackrel{def}{=} \frac{\delta(\sqrt{-g}\,L_M)}{\delta h^i_{.\lambda}} = \frac{\partial(\sqrt{-g}\,L_M)}{\partial h^i_{.\lambda}} - \frac{\partial}{\partial x^{\mu}}(\frac{\partial(\sqrt{-g}\,L_M)}{\partial h^i_{.\lambda,\mu}})$$

$$= h^{\alpha}_i\,(\sqrt{-g}\,L_M\,\delta^{\lambda}_{\alpha} - \frac{\partial(\sqrt{-g}\,L_M)}{\partial \psi_{,\lambda}}\psi_{,\alpha} - \frac{\partial(\sqrt{-g}\,L_M)}{\partial h^i_{.\mu,\lambda}}h^i_{.\mu,\alpha}) \qquad (43)$$

In Eq.(43) the relation represented by Eq.(41) is used. The definition of Eq.(43) is equivalent to the definition

$$\sqrt{-g}\,T^{\mu\nu}_{(G)} \stackrel{def}{=} 2\frac{\delta(\sqrt{-g}\,L_G)}{\delta g_{\mu\nu}}\,. \quad T^{.\lambda}_{(M)i}$$ is a tensor with mixed indexes, there exists the relation:

$T^{.\lambda}_{(M)\alpha} = h^i_{.\alpha} T^{.\lambda}_{(M)i}$. Some scholars define the energy-momentum tensor density of pure gravitational field as

$$t^{\lambda}_{(G)\alpha} \stackrel{def}{=} \sqrt{-g} L_G \delta^{\lambda}_{\alpha} - \frac{\partial(\sqrt{-g} L_G)}{\partial h^i_{.\mu,\lambda}} h^i_{.\mu,\alpha} - \frac{\partial(\sqrt{-g} L_G)}{\partial h^i_{.\mu,\lambda\sigma}} h^i_{.\mu,\sigma\alpha}$$
$$+ \frac{\partial}{\partial x^{\sigma}}(\frac{\partial(\sqrt{-g} L_G)}{\partial h^i_{.\mu,\lambda\sigma}}) h^i_{.\mu,\sigma\alpha} \quad (44)$$

Then the conservation laws of energy-momentum tensor density for a gravitational system may be expressed by

$$\frac{\partial}{\partial x^{\lambda}}(\sqrt{-g} T^{\lambda}_{(M)\alpha} + \sqrt{-g} t^{\lambda}_{(G)\alpha}) = 0 \quad (45)$$

Which is equivalent to the Einstein's conservation laws. But the quantity $t^{\lambda}_{(G)\alpha}$ is not tensor, that Eq.(45) lacks the invariant character required by the principles of general relativity is its serious defects. There is another definition, some other scholars define the energy-momentum tensor density of pure gravitational field as

$$\sqrt{-g} T^{\lambda}_{(G)i} \stackrel{def}{=} \frac{\delta(\sqrt{-g} L_G)}{\delta h^i_{.\lambda}}$$
$$= \frac{\partial(\sqrt{-g} L_G)}{\partial h^i_{.\lambda}} - \frac{\partial}{\partial x^{\mu}}(\frac{\partial(\sqrt{-g} L_G)}{\partial h^i_{.\lambda,\mu}}) + \frac{\partial^2}{\partial x^{\mu} \partial x^{\sigma}}(\frac{\partial(\sqrt{-g} L_G)}{\partial h^i_{.\lambda,\mu\sigma}})$$
$$= h^{\alpha}_i (\sqrt{-g} L_G \delta^{\lambda}_{\alpha} - \frac{\partial(\sqrt{-g} L_G)}{\partial h^i_{.\mu,\lambda}} h^i_{.\mu,\alpha} - \frac{\partial(\sqrt{-g} L_G)}{\partial h^i_{.\mu,\lambda\sigma}} h^i_{.\mu,\sigma\alpha} \quad (46)$$
$$+ \frac{\partial}{\partial x^{\sigma}}(\frac{\partial(\sqrt{-g} L_G)}{\partial h^i_{.\mu,\lambda\sigma}}) h^i_{.\mu,\alpha} + \frac{\partial^2}{\partial x^{\mu} \partial x^{\sigma}}(\frac{\partial(\sqrt{-g} L_G)}{\partial h^i_{.\lambda,\mu\sigma}} h^i_{.\alpha}))$$

$T^{.\lambda}_{(G)i}$ is also a tensor with mixed indexes, and we have: $T^{.\lambda}_{(G)\alpha} = h^i_{.\alpha} T^{.\lambda}_{(G)i}$. Thus the conservation laws of energy-momentum tensor density for a gravitational system may be expressed also by

$$\frac{\partial}{\partial x^{\lambda}}(\sqrt{-g} T^{\lambda}_{(M)\alpha} + \sqrt{-g} T^{\lambda}_{(G)\alpha}) = 0$$

which is exactly the Lorentz and Levi-Civita's conservation laws, *i.e.* Eq.(2). Eq.(38) tell us that

$$T^{\lambda}_{(M)\alpha} + T^{\lambda}_{(G)\alpha} = 0$$

which is just Eq.(3). It must indicate that $T^{.\lambda}_{(G)\alpha}$ relates with $t^{\lambda}_{(G)\alpha}$ by (compare Eq.(44) with Eq.(46))

$$\sqrt{-g}\, T^{\lambda}_{(G)\alpha} = \sqrt{-g}\, t^{\lambda}_{(G)\alpha} + \frac{\partial^2}{\partial x^{\mu} \partial x^{\sigma}} \left( \frac{\partial(\sqrt{-g}\, L_G)}{\partial h^i_{.\lambda,\mu\sigma}} h^i_{.\alpha} \right) \qquad (47)$$

Consequently $\frac{\partial}{\partial x^{\lambda}}(\sqrt{-g}\, T^{\lambda}_{(M)\alpha} + \sqrt{-g}\, T^{\lambda}_{(G)\alpha}) = \frac{\partial}{\partial x^{\lambda}}(\sqrt{-g}\, T^{\lambda}_{(M)\alpha} + \sqrt{-g}\, t^{\lambda}_{(G)\alpha}) = 0$, because Eq.(29), $\frac{\partial^3}{\partial x^{\lambda} \partial x^{\mu} \partial x^{\sigma}} \left( \frac{\partial(\sqrt{-g}\, L_G)}{\partial h^i_{.\lambda,\mu\sigma}} h^i_{.\alpha} \right) = 0.$ These relations tell us that the Lorentz and Levi-Civita's conservation laws is conneting intimately with the Einstein's conservation laws.

Einstein did not agree with Eq.(3) [9], because he believed that the relation expressed by Eq. (3) should make the energy-momentum of a material system, being $T_{(M)\mu\nu} \neq 0$ in the initial state, to $T_{(M)\mu\nu} \to 0$ spontaneously. By using Boltzmann's relation S=k ln N, we have shown that this view is incorrect [1,10]. An important debate was evoked about the definitions of energy-momentum tensor density for gravitational field and the related conservation laws in 1917-1918 [9]; Einstein is on the one side of that debate, his opponents are Levi-Civita and others. This debate had not reached unanimity, but because Einstein enjoyed great prestige among academic circles and many scholars followed him, therefore the definition Eq.(44) and the Einstein's conservation laws have become the prevalent views now in the gravitational theory. The author hold that, as the Lorentz and Levi-Civita's conservation laws being to connect intimately with the Einstein's conservation laws, these two conservation laws are all well worth to consider carefully. Which law is correct on physical side ? This question can answer only by experimental and observational tests. To affirm subjectively a law is not suitable. In the last few years the author have thoroughly studied Lorentz and Levi-Civita's conservation laws and found that these conservation laws have abundant physical contents [1,10-13] which can be tested via experiments or observations. In these respects, the Lorentz and Levi-Civita's conservation laws will undoubtedly be used as one of important theoretical foundations to establish a new cosmology.

From Eq.(34'), i.e. $\frac{\delta(\sqrt{-g}\, L_G)}{\delta h^i_{.\mu}} = -\frac{\delta(\sqrt{-g}\, L_M)}{\delta h^i_{.\mu}}$ and the definition $\sqrt{-g}\, T^{.\mu}_{(M)i} = \frac{\delta(\sqrt{-g}\, L_M)}{\delta h^i_{.\mu}}$

the gravitational field equations take the form:

$$\frac{\delta(\sqrt{-g}\,L_G)}{\delta h^i_{\cdot\mu}} = -\sqrt{-g}\,T^{\cdot\mu}_{(M)i} \qquad (48)$$

For $L_G(x) = \frac{1}{16\pi G} R(x)$, $\frac{\delta(\sqrt{-g}\,L_G)}{\delta h^i_{\cdot\mu}} = \frac{1}{8\pi G}\sqrt{-g}\,(R^{\cdot\mu}_i - \frac{1}{2} g^{\cdot\mu}_i R)$, we get Einstein gravitational field equations without cosmological constant :  $R^{\mu\nu} - \frac{1}{2} g^{\mu\nu} R = -8\pi G\,T^{\mu\nu}_{(M)}$ ;

For $\sqrt{-g}\,L_G(x) = \frac{\sqrt{-g}}{16\pi G}[R(x) + 2\lambda]$, $\frac{\delta(\sqrt{-g}\,L_G)}{\delta h^i_{\cdot\mu}} = \frac{1}{8\pi G}\sqrt{-g}\,(R^{\cdot\mu}_i - \frac{1}{2} g^{\cdot\mu}_i R - \lambda g^{\cdot\mu}_i)$, we get Einstein gravitational field equations with cosmological constant :

$$R^{\mu\nu} - \frac{1}{2} g^{\mu\nu} R - \lambda g^{\mu\nu} = -8\pi G\,T^{\mu\nu}_{(M)} \; ;$$

For $\sqrt{-g}\,L_G(x) = \frac{\sqrt{-g}}{16\pi G}[R(x) + 2\lambda + 2D(t)]$,

$\frac{\delta(\sqrt{-g}\,L_G)}{\delta h^i_{\cdot\mu}} = \frac{1}{8\pi G}\sqrt{-g}\,(R^{\cdot\mu}_i - \frac{1}{2} g^{\cdot\mu}_i R - \lambda g^{\cdot\mu}_i - D g^{\cdot\mu}_i)$, we get the modified Einstein's field equations in the new theory of cosmology established at Ref.[1] :

$$R^{\mu\nu} - \frac{1}{2} g^{\mu\nu} R - \lambda g^{\mu\nu} - D^{\mu\nu} = -8\pi G\,T^{\mu\nu}_{(M)} \quad (D^{\mu\nu} \stackrel{def}{=} D g^{\mu\nu}).$$

This equation is just Eq.(1).

It is evident that all of these gravitational field equations are fully rational and well worth a more careful examination. Which gravitational field equations is correct ? This question can only be answered by experimental and observational tests. To affirm subjectively a gravitational field equations is not suitable. Anyhow, the modified Einstein's field equations *i.e.* Eq.(1) will undoubtedly be used as one of important theoretical foundations to establish a new cosmology.

**4. Some comparisons between the standard big bang cosmology and the new cosmology established in Ref.[1]**

The detailed content of the new cosmology have been presented in Ref.[1], we shall not repeat to write it in this paper; we shall make only some comparisons between the standard big bang cosmology (SBBC) and the new cosmology (NC) established in Ref.[1].

The common or similar character of SBBC and NC are:

1), The universe is assumed to be spatially homogeneous and isotropic [1] ( this assumption is called cosmological principle), so the universe has the Robertson-Walker metric [5]

$$d\tau^2 = -dt^2 + a(t)^2 \{ \frac{dr^2}{1-kr^2} + r^2 d\theta^2 + r^2 \sin^2\theta d\phi^2 \}$$, and the energy-momentum tensor

of the matter field should take the form of ideal fluid [5] $T^{\mu\nu}_{(M)} = (\rho_M + p_M) u^\mu u^\nu + p_M g^{\mu\nu}$.

2), The universe is expanding.

3), These two cosmologies all could explain the microwave radiation background [1,5]; in SBBC it is believed that the microwave radiation background is just the left-over radiation from the big bang; in NB it is supposed that the universe might have taken place a change from the radiation-dominated era to the matter-dominated era which is similar with SBBC, and it is considered that the microwave radiation background is just the left-over radiation from this change.

The different character of SBBC and NC are:

1), In SBBC it is believed that the universe has a beginning state called the big bang, but in NB it is believed that the big bang might never have occurred since NB maintains that the universe is without a beginning and without an end and considers that the state $\rho_M \to \infty$ might never have existed [1].

2), In NB there is the possibility that the energy of matter field might originate from the gravitational field [1], but SBBC does not study the origin of the matter field's energy, it is assumed that the total energy of matter fields (including the inflation field) has existed since the big bang.

3), In SBBC the observed abundances of light nuclei in the universe are explained as the result of nucleon-synthesis taking place in the early universe at a temperature of about $10^9$ ($^0K$). In NB although the observed abundances of light nuclei in the universe could be explained with the same reason as SBBC, but it also could be explained that the light nuclei in the universe are formed from hydrogen nuclei in the interiors of stars [1].

4), NB affirms that $\lambda g^{\mu\nu}$ and $D^{\mu\nu}$ are two quantities to represent the pure gravitational field and does not consider them as the quantities to represent matter field in essence, $\lambda g^{\mu\nu}$ and $D^{\mu\nu}$ could be interpreted as dark energy and a main part of dark matter respectively; but in SBBC although $\lambda g^{\mu\nu}$ could be interpreted as dark energy also, yet it is always considered to be a part of matter field; especially there is no quantity $D^{\mu\nu}$ in SBBC, the dark matter is looked entirely as some kinds of material matter [1].

From the above comparisons it is evident that the range of study for NB is greater than that for SBBC; moreover, through studing NB several experimentst and observations to test whether SBBC or NB is correct are advanced.As many new evidences of observations [14, 15, 16] have brought out some crucial weaknesses of SBBC. It is necessary to introduce new concepts and new theories, so I believe that to study NB is significant.